\documentclass[pdftex,twocolumn,epjc3]{svjour3}          

\RequirePackage[T1]{fontenc}

\RequirePackage{graphicx}
\RequirePackage{mathptmx}      
\RequirePackage{flushend}
\RequirePackage[numbers,sort&compress]{natbib}
\RequirePackage[colorlinks,citecolor=blue,urlcolor=blue,linkcolor=blue]{hyperref}

\usepackage{amssymb}

\begin{document}

\title{The seesaw path to leptonic CP violation}

\author{A. Caputo\thanksref{e1,addr1,addr2}, P. Hernandez \thanksref{e2,addr1,addr2}, M. Kekic \thanksref{e3,addr1}, J. L\'opez-Pav\'on \thanksref{e4,addr2} 
       \and
        J. Salvado\thanksref{e5,addr1} 
}

\thankstext{e1}{ andrea.caputo@cern.ch}
\thankstext{e2}{ m.pilar.hernandez@uv.es}
\thankstext{e3}{ marija.kekic@ific.uv.es}
\thankstext{e4}{jacobo.lopez.pavon@cern.ch}
\thankstext{e5}{jordi.salvado@ific.uv.es}
\institute{Instituto de F\'{\i}sica Corpuscular, Universidad de Valencia and CSIC, \\
 Edificio Institutos Investigaci\'on, Catedr\'atico Jos\'e Beltr\'an 2, 46980 Spain. \label{addr1}
 \and
CERN, Theoretical Physics Department, Geneva, Switzerland.  \label{addr2}
}


\maketitle

\begin{abstract}
Future experiments such as SHiP and high-intensity $e^+ e^-$ colliders will have a superb sensitivity to 
heavy Majorana neutrinos with masses below $M_Z$. We show that the measurement 
of the mixing to electrons and muons of one such state could imply the discovery of leptonic 
CP violation in the context of seesaw models. 
We quantify in the minimal model the CP discovery potential of these future experiments, and demonstrate
that a 5$\sigma$ CL discovery of leptonic CP violation would be possible in a very significant fraction of 
parameter space. 
\end{abstract}

\section{Introduction}

The simplest extension of the Standard Model that can accommodate naturally light neutrino masses is the well-known seesaw
model \cite{seesaw}, where at least two  singlet  Majorana fermions, $N_R^i$, are added and can couple to the lepton doublets via
a Higgs-Yukawa coupling. 

The Lagrangian of the model is given by:
   \begin{eqnarray}
{\cal L} = {\cal L}_{SM}- \sum_{\alpha,i} \bar L^\alpha Y^{\alpha i} \tilde\Phi N^i_R - \sum_{i,j=1}^2 {1\over 2} \bar{N}^{ic}_R M^{ij} N_R^j+ h.c., \nonumber
\label{eq:lag}
\end{eqnarray}
where $Y$ is a $3\times 2$ complex matrix and $M$ is a two-dimensional  complex symmetric matrix. 

While the Majorana mass scale has been traditionally assumed  to be very large so that the light 
neutrino masses can be explained with yukawa couplings of ${\mathcal O}(1)$,  the absence of any indication of new physics that solves the
 hierarchy problem suggests that a more natural option would be to assume a Majorana scale of the order of the electroweak scale. The lepton flavour puzzle 
would not be qualitatively different to the quark one, at least in what concerns the spread in the yukawas, since the neutrino yukawas
would need to be only slightly smaller than the electron and light-quark ones. 
From a phenomenological point of view this is of course a very interesting possibility since the heavy Majorana neutrino states might be searched for in fixed-target experiments and at  colliders. 

In this letter we explore the opportunities that this opens for the discovery of leptonic CP violation. In particular we concentrate on the observables
that could be provided by direct searches of these heavy neutrino states, more concretely their mixings to electrons and muons. It is well-known 
that in seesaw models there is a strong correlation between the mixings of the heavy states and the light and heavy neutrino spectrum. In particular for just one neutrino species/flavour,  the mixing of the heavy state is  fixed to be
\begin{eqnarray}
|U_{\alpha h}|^2 = {m_l\over M_h} ,
\label{eq:naive}
\end{eqnarray}
  where $m_l$ is the light neutrino mass and $M_h$ is the heavy one. This is the naive seesaw scaling which implies that the mixing of heavy states is highly suppressed when $M_h \gg m_l$.
  
In the case of more families the correlation still exists but it is not so strong. In particular, for the minimal model with two heavy neutrinos, using the Casas-Ibarra parametrization of the model \cite{Casas:2001sr}, the mixing of the heavy neutrinos is given in all generality by 
\begin{eqnarray}
U_{\alpha h} = i U_{\rm PMNS} \sqrt{m_l}~ P_{NO} ~R^\dagger(z) M^{-1/2},
\end{eqnarray} 
where $m_l$ is the diagonal matrix of the light neutrino masses (note that the lightest neutrino is massless) and $U_{\rm PMNS}$ is the PMNS matrix in the standard parametrization, which depends only on two CP violating phases \cite{Cabibbo:1977nk,Bilenky:1980cx,Schechter:1980gr}.  
$P_{\rm NO}$ is a  $3\times 2$ matrix that depends on the neutrino ordering (NH, IH)
\begin{eqnarray}
 P_{NH} =\left(\begin{array}{l} {\mathbf 0}\\ {\mathbf I}\end{array}\right),~ P_{IH} =\left(\begin{array}{l} {\mathbf I}\\ {\mathbf 0}\end{array}\right),
 \end{eqnarray}
where ${\mathbf I}$ is the $2\times 2$ identity matrix and ${\mathbf 0}=(0,0)$ \cite{Donini:2012tt}. The unknown parameters are 
$M={\rm Diag}(M_1,M_2)$, the diagonal matrix of the heavy neutrino masses, and $R$, a  two-dimensional complex orthogonal matrix,  that depends generically on one complex angle, $z=\theta + i \gamma$. The entries in this matrix can be very large and therefore the naive seesaw scaling of eq.~(\ref{eq:naive}) is not generically satisfied in the case of more than one family. In this work we will fix the known oscillation parameters to their best fit values as taken from \cite{Esteban:2016qun}. 

In \cite{Hernandez:2016kel}, we pointed out that in the region of large mixings  the ratio of mixings to electron and muons of the heavy states
are essentially fixed by the PMNS matrix, and  is therefore very sensitive to its unknown CP phases. In this paper, our aim is to quantify the CP discovery potential of such measurement.  Interestingly this measurement provides an example of a CP-conserving 
observable that is sensitive to leptonic CP-violation.

It should be noted that the results below will also apply to the model with three neutrinos, where one of them is sufficiently decoupled, as in the $\nu$MSM \cite{nuMSM}. 

\section{Heavy neutrino mixings versus leptonic CP phases}

Given the upper bound on the light neutrino masses of ${\mathcal O}(1)$eV, the naive seesaw scaling formula of eq.~(\ref{eq:naive}) implies that the mixings of the heavy states are of $|U_{\alpha h}|^2 \simeq {\mathcal O}\left({10^{-9} GeV \over M }\right)$, a value at the limit of sensitivity of SHiP \cite{Anelli:2015pba} in the GeV range or FCC-e in the ${\mathcal O}(10)$GeV range \cite{Blondel:2014bra}.  This implies that in most of the sensitivity range of SHiP and other collider experiments the entries of $R$ need to be largish (ie. $\gamma \gtrsim 1$), and in this case the dependence of the mixings on the unknown parameters simplifies greatly. Indeed, a perturbative expansion in the small parameters 
 \begin{eqnarray}
\mathcal{O}\left(\epsilon \right): r\equiv\sqrt{\frac{\Delta m^2_{sol}}{\Delta m^2_{atm}}}\sim  \theta_{13}  \sim e^{-{\gamma\over 2}},
\end{eqnarray}
 shows that the ratio of electron/muon mixings does not depend on the complex angle $\gamma, \theta$,  nor on the masses of the heavy states, and only depends on the  the mixing angles and CP phases of the PMNS matrix \cite{Hernandez:2016kel}. Defining $A\equiv \frac{e^{2\gamma}\sqrt{\Delta m^2_{atm}}}{4}$, the result for the inverted ordering (IH) is : 
\begin{eqnarray}
|U_{ei}|^2 M_i  &\simeq& A \Big[ (1 + \sin\phi_1\sin2\theta_{12})(1-\theta_{13}^2) \nonumber\\
& +& {1\over 2} r^2 s_{12} (c_{12} \sin\phi_1+s_{12})+{\mathcal O}(\epsilon^3)\Big] ,\nonumber\\
|U_{\mu i}|^2 M_i  &\simeq&A \Big[ \left
(1 - \sin\phi_1\sin2\theta_{12} \left(1+ {1\over 4}r^2 \right)+ {1\over 2} r^2 c^2_{12} \right) c_{23}^2 \nonumber\\
&&+ \theta_{13} (\cos \phi_1 \sin \delta - \sin\phi_1 \cos 2 \theta_{12} \cos\delta) \sin 2 \theta_{23}\nonumber\\
&&+ \theta_{13}^2 (1+\sin\phi_1 \sin 2 \theta_{12}) s_{23}^2  +{\mathcal O}(\epsilon^3)\Big],\nonumber\\
\label{eq:uihanal}
\end{eqnarray}
while for the normal one (NH):
\begin{eqnarray}
|U_{ei}|^2 M_i  &\simeq& A\Big[  r s_{12}^2  -2  \sqrt{r} \theta_{13} \sin(\delta+\phi_1) s
_{12}+ \theta_{13}^2 +{\mathcal O}(\epsilon^{5/2})\Big] ,\nonumber\\
|U_{\mu i}|^2 M_i  &\simeq&A\Big[ s_{23}^2  - \sqrt{r}~  c_{12} \sin\phi_1 \sin 2\theta_{23}  + r c_{12}^2 c_{23}^2 \nonumber\\
&&+ 2  \sqrt{r} ~\theta_{13} \sin(\phi_1+\delta) s_{12} s_{23}^2 - \theta_{13}^2 s_{23}^2 +{\mathcal O}(\epsilon^{5/2})\Big].\nonumber\\
\label{eq:unhanal}
\end{eqnarray}

On the other hand, the parameter $\gamma$ and the Majorana masses determine the global size of the mixings, and therefore the statistical uncertainty in  the measurement of the ratio of mixings. 

However, values of the heavy mixings much larger that the naive seesaw scaling in eq.~(\ref{eq:naive}) imply generically large one loop corrections to neutrino masses \cite{LopezPavon:2012zg} except in the region where an approximate lepton number symmetry holds \cite{u1}, that implies highly degenerate neutrinos, besides large $\gamma$. 

The question we want to quantify is what is the range of the PMNS CP-phase parameter space, that is the rectangle $\delta, \phi_1 \in [0, 2 \pi]$, in which leptonic CP violation can be discovered via the measurement of the electron and muon mixings of one of the heavy neutrinos. In order words for which values of the true parameters $(\delta, \phi_1)$ can these mixings  be distinguished from those at the CP conserving points $(0,0)$, $(0,\pi)$, $(\pi,0)$, $(\pi,\pi)$.

\section{Discovery of Majorana neutrinos  }

We consider two future experiments that have the potential to discover sterile neutrinos in complementary regions of masses: SHiP \cite{Anelli:2015pba} and FCC-ee \cite{Blondel:2014bra}.

\subsection{The SHiP experiment}

The SHiP experiment \cite{Anelli:2015pba} will search for the heavy neutrinos of the seesaw model in charmed and b meson decays. Neutrinos with masses below charmed and b meson masses could be discovered, and the best sensitivity is expected for masses around 1 GeV. In order to carry out our study we would ideally need an estimate
of the uncertainty in the determination of the mixing angles of the sterile neutrinos  to electrons and muons as a function of those mixings and the heavy neutrino masses. Such information has not been published yet by the collaboration. The only publicly available result is the sensitivity reach (corresponding to 2 expected events in a 5 year run with 0.1 estimated background events) on the plane $U^2 \equiv \sum_\alpha |U_{\alpha i}|^2$ vs $M_i$ for five scenarios, where  the contribution of electron and muon channels  is combined, 
and only the channels with two charged particles in the final state are considered.  These scenarios also correspond to two highly degenerate neutrinos with equal mixings that contribute equally to the signal. Since we are interested in a more general case, where neutrinos might not be degenerate, we consider the contribution of just one of the states, so we will assume that at the sensitivity limit the number of events is one instead of two. 

The number of events in the charged-current two body decays $N\rightarrow l_\alpha^\pm H^\mp$, for different hadrons $H=\pi,\rho,...$,  is expected to scale as 
\begin{eqnarray}
N_\alpha(M_i) \propto |U_{\alpha i}|^2 U^2,
\label{eq:scal}
\end{eqnarray}
because $U^2$ controls the number of charm decays into sterile neutrinos, while $|U_{\alpha i}|^2$ controls the fraction of those that decay to the lepton flavour $\alpha$ via a charged current within the detector length (which is assumed much smaller than the decay length). Note that in the charged-current three body decays 
$N\rightarrow l^-_\alpha l^+_\beta \nu_\beta$ with $\alpha\neq \beta$, the contributions from the charged current muon  and electron mixings cannot be distinguished, because the $N$'s are Majorana. Also those decays with $\alpha =\beta$ cannot be distinguished  from the corresponding neutral-current process. These three body decays are however subleading in the sensitivity regime of SHiP so we will consider only the information from the two body decays. 
 
\begin{figure}[h]
 \begin{center}
\includegraphics[scale=0.6]{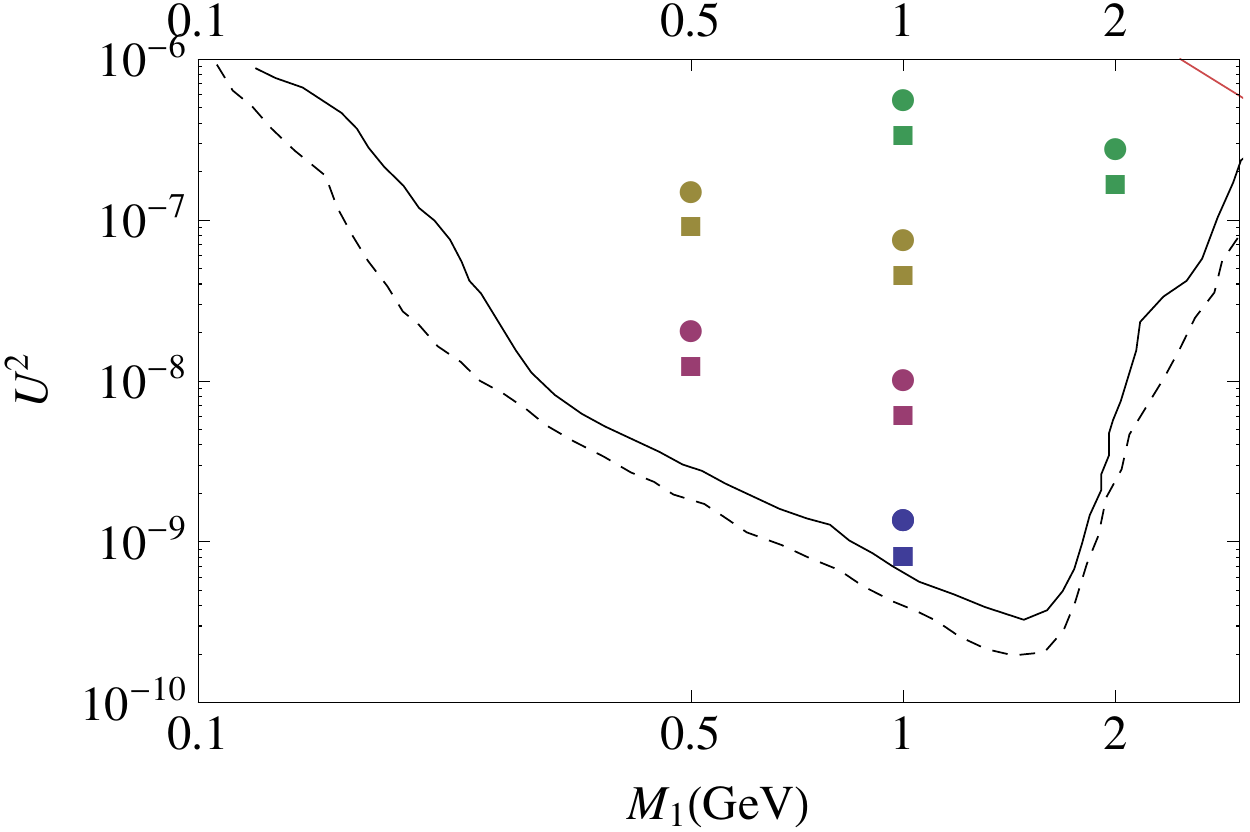} 
\caption{\label{fig:ship} Lines corresponding to two expected events for scenarios II (solid) and IV (dashed) in \cite{Anelli:2015pba}. The indicated points
correspond to the CP conserving case and $\gamma=2,3,4,5$ in ascending order for IH (circles) and NH (squares). The red line in the top right corner 
corresponds to one-loop
corrections to neutrino masses being of the same order as the atmospheric splitting for $ \Delta M/M_1 = 1$. }
\end{center}
\end{figure}

Among the five scenarios studied in \cite{Anelli:2015pba,shipnote1}, there are however two of them where either the mixing to electrons or the mixing to muons dominates by a large factor. In Fig.~\ref{fig:ship} we reproduce the SHiP sensitivity curves for these electron-dominated (IV) and muon-dominated (II) scenarios. As an illustration, we also indicate in Fig.~\ref{fig:ship} the mixings corresponding to different values of the Casas-Ibarra parameter $\gamma$. Stringent bounds exist from direct searches. We do not include them in the plot because they are flavour dependent and cannot be easily interpreted in terms of $U^2$. They restrict severely the region of masses below $0.5$ GeV and cut slightly the range of largest mixings. We will therefore not consider this region of parameter space. In the figure we also indicate the 
line that corresponds to one-loop corrections to neutrino mass differences being of the same order as the atmospheric mass splitting, for a level of degeneracy 
$\Delta M/M_1 =1$. No degeneracy is therefore required in this range of masses to reach mixings up to $U^2 \sim 10^{-6}$ without
fine-tunning. 

In order to estimate the number of electron and muon events as a function of the Casas-Ibarra parameters, we assume that the total number of events (with two charged particles in the final state) is one at the sensitivity limit curve in the two scenarios, and we compute the contribution to this total of the electron and muon two-body decay channels. Note that for $M\sim 1$GeV the three-body decays are subleading. At each mass, the number of electron and muon events at different value of the mixings is obtained by scaling this number according to  eq.~(\ref{eq:scal}). 
 
Besides the number of electron and muon events we  assume  that the mass of the heavy neutrino is measured, which should be easy in these two body channels. Again the uncertainty in the determination of the mass as a function of the mass and mixing has not yet been presented by the SHiP collaboration. A realistic error is much harder to estimate in this case, since it involves reconstructing a peak in the invariant mass. We will therefore assume a fixed relative error of $1\%$.

\subsection{FCC-ee collider }

For masses above 2 GeV, the heavy neutrinos can be searched for in high luminosity colliders such as the future $e^+ e^-$ circular collider (FCC-ee) \cite{Blondel:2014bra,Abada:2014cca,Antusch:2015mia} that can provide
$10^{12}-10^{13}$ $Z$ boson decays at rest per year. Sterile neutrinos with masses $m_N \leq M_Z$ can be produced in the decay $Z\rightarrow N \nu$, and leave a very characteristic signal of a displaced vertex produced by the decay of the long-lived heavy neutrino. A simplified estimate of the number of events to visible leptonic channels within the detector can be obtained as follows:
\begin{eqnarray}
N_{\rm total} = N_Z ~Ê{\rm BR}(Z \rightarrow N \nu)  {\rm BR}(N \rightarrow {\rm leptonic}) \nonumber\\
\times  \left(e^{- l_{\rm min} /\gamma_L c \tau_N}-e^{- l_{\rm max} /\gamma_L c \tau_N}\right) ,
\end{eqnarray}
where $N_Z$ is the number of $Z$ decays, $\gamma_L={1\over 2}\left({M_Z\over  M_1} + {M_1 \over  M_Z}\right)$ is the heavy neutrino Lorentz boost gamma factor and $\tau_N$ is its lifetime.  $l_{\rm min}$ and $l_{\rm max}$ are the minimum and maximum displacement of the secondary vertex that can be measured. 
The $Z\rightarrow N\nu$ branching ratio has been computed in \cite{Dittmar:1989yg} to be
\begin{eqnarray}
{\rm BR}(Z\rightarrow N_i\nu) = 2 U^2 {\rm BR}(Z \rightarrow \nu {\bar \nu}) \left(1-{M_i^2\over M_Z^2}\right)^2 \left(1+{1 \over 2} {M_i^2\over M_Z^2}\right),\nonumber\\
\end{eqnarray} 
where ${\rm BR}(Z \rightarrow \nu\bar{\nu})$ corresponds to one family. 

The partial width to the  leptonic channels can be found in many references, see for example \cite{Atre:2009rg}, while the total width requires an estimate of the hadronic width. 
While in the SHiP case, exclusive hadronic channels where considered, for the heavier masses relevant at FCC-ee,  the inclusive hadronic decay width is approximated by the parton model \cite{Gronau:1984ct}.  This is sufficient for our purposes. Note however that a precise determination of this width would be possible, applying similar methods to those used in $\tau$ decays \cite{Braaten:1988hc}. 

\begin{figure}[h]
 \begin{center}
\includegraphics[scale=0.6]{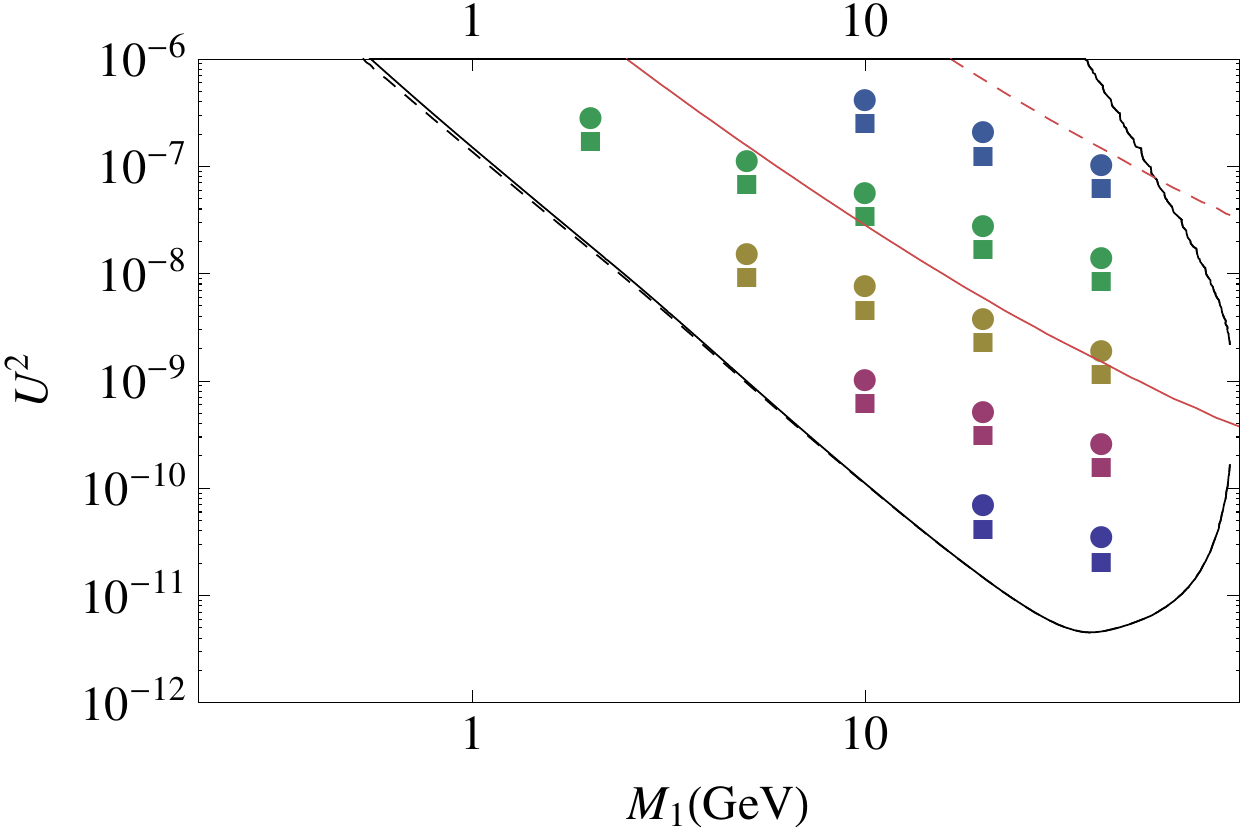} 
\caption{\label{fig:fcc} Lines corresponding to two expected events for scenarios II (solid) and IV (dashed) in \cite{shipnote1}. The indicated points
correspond to the CP conserving case and $\gamma=2,3,4,5,6$ in ascending order for IH (circles) and NH (squares). The red lines in the top right corner 
corresponds to one-loop
corrections to neutrino masses being of the same order as the atmospheric splitting for $\Delta M/M_1 = 1$ (solid) and  $\Delta M/M_1 = 0.01$ (dashed). }
\end{center}
\end{figure}

  In the recent analysis of \cite{shipnote1}, a FCC-ee configuration with $10^{13}$ Z decays per year and an inner tracker that can resolve a displaced vertex at a distance between $l_{\rm min} = 0.1$ mm and $l_{\rm max} = 5$ m  
 has been assumed in the context of the same two scenarios above. The sensitivity reach using only leptonic channels and assuming $100\%$ efficiency and zero background  is reproduced in Fig.~\ref{fig:fcc}, which agrees reasonably well with the results of  \cite{shipnote1}. The regions of mixings above the red lines would imply fine-tunning between the tree level and one-loop contributions to the light neutrino masses for a level of degeneracy of $ \Delta M/M_1 = 1$ (solid) and 
 $\Delta M/M_1 = 0.01$ (dashed). 

  Since we are interested in measuring separately the mixings to electrons and muons, we will separate the distinguishable leptonic channels $l_\alpha l_\beta=ee, e\mu, \mu\mu$. Note that each of these channels depends on several mixings, either because they get a contribution from neutral currents in the case of $\beta=\alpha$ or because  the channels where the leading lepton (ie. that coupled to the $N_i$) is $\alpha$ or $\beta\neq \alpha$ are indistinguishable. We  also consider the inclusive hadronic ones $N\rightarrow l_\alpha + q +\bar{q}'$ that depend only on the mixing $|U_{\alpha i}|$.  
 Finally we assume an angular acceptance of $\sim 80\%$ (ie we require the $N$ direction is at least $15^\circ$ off  the beam axis) and include just statistical uncertainties. 
 
\section{CP discovery potential }

In order to quantify the discovery CP potential we consider that SHiP or FCC-ee will measure the number of electron and muon events in the decay of one of the heavy neutrino states  (without loss of generality we assume to be that with mass $M_1$), estimated as explained in the previous section. We will only consider statistical errors. 

 The test statistics (TS) for leptonic CP violation is then defined as follows:
\begin{eqnarray}
\Delta \chi^2
&\equiv&-2\sum_{\alpha = {\rm channel}} N^{\rm true}_\alpha -N^{CP}_{\alpha}+N^{\rm true}_\alpha \log\left( {N^{\rm CP}_\alpha\over N^{\rm true}_{\alpha} }\right)
\nonumber\\
&+&  \left( { M_1 - {M^{\rm min}_1} \over \Delta M_1}\right)^2.
\end{eqnarray}
where $N_\alpha^{\rm true} = N_\alpha(\delta, \phi_1,M_1,\gamma,\theta )$ is the number of events for the true model parameters,  and $N_\alpha^{CP} = N_{\alpha}(CP, \gamma^{\rm min}, \theta^{\rm min}, M_1^{\rm min})$ is the number of events for the CP-conserving test hypothesis that minimizes $\Delta \chi^2$ among the four CP conserving phase choices $CP=(0/\pi,0/\pi)$ and 
over the unknown test parameters. $\Delta M_1$ is the uncertainty in the mass, which is assumed to be 
$1\%$. 

\begin{figure}[h]
 \begin{center}
\includegraphics[scale=0.4]{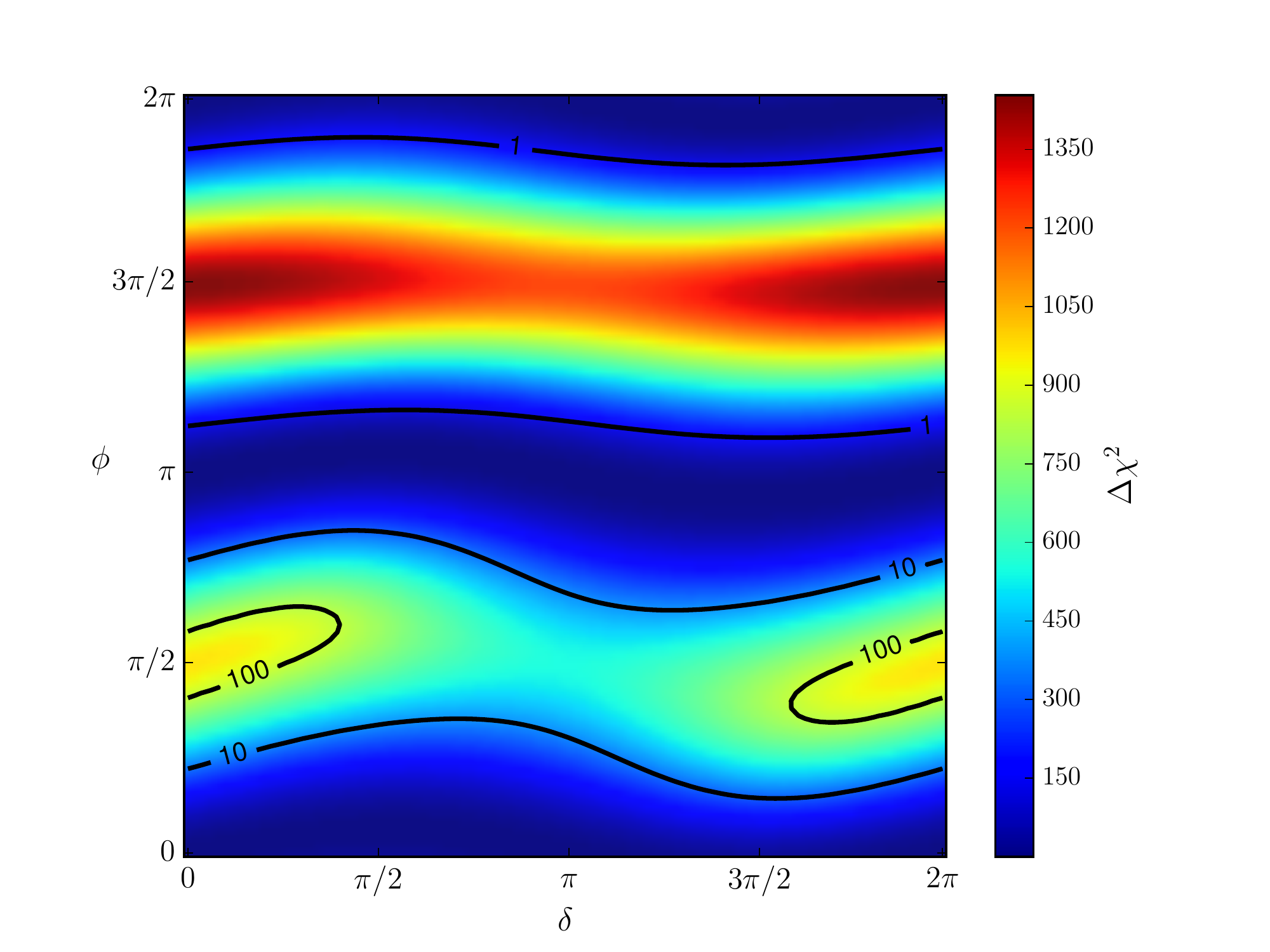}  
\includegraphics[scale=0.4]{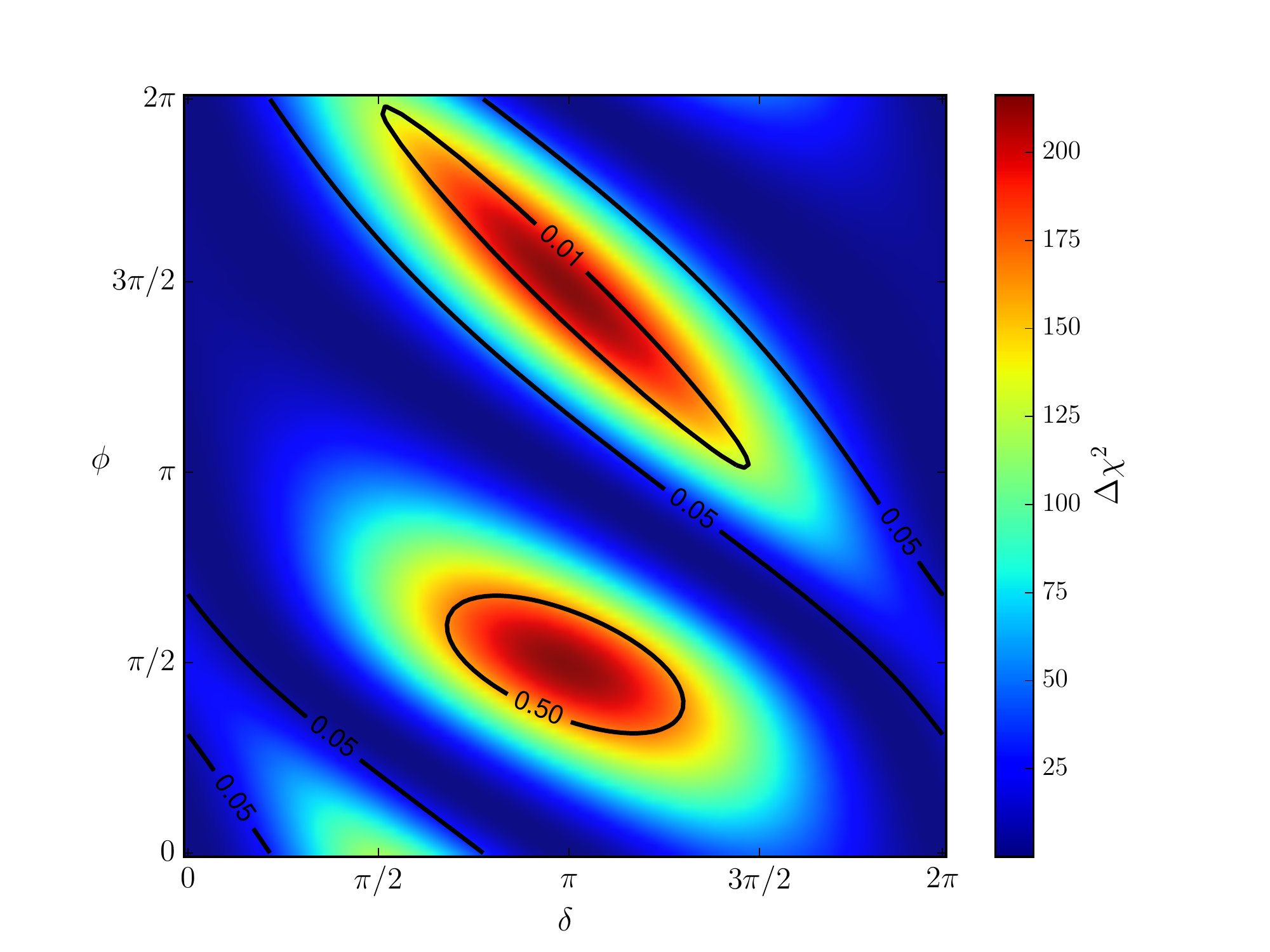}  
\caption{\label{fig:contours} Surface plots of $\Delta \chi^2$ for $ (\gamma,\theta, M_1) = (3.5,0, 1$ GeV) as a function of the true $(\delta, \phi_1)$ of IH (up) and NH (down) for SHiP. The lines correspond to the constant value indicated of the ratio of electron-to-muon mixing as obtained from the analytical formulae of eqs.~(\ref{eq:uihanal}-\ref{eq:unhanal}).}
\end{center}
\end{figure}

Surface plots for $\Delta \chi^2$ are shown in Fig.~\ref{fig:contours} for SHiP and the true parameters $\{\gamma,\theta,M_1\} = \{3.5, 0, 1 {\rm GeV}\}$. The basic features of these contours can be understood analytically, as shown by the superimposed 
lines corresponding to a constant electron-to-muon mixing ratio, as obtained from the analytical formulae in eqs.~(\ref{eq:uihanal}) and (\ref{eq:unhanal}).

\begin{figure}[h]
 \begin{center}
\includegraphics[scale=0.4]{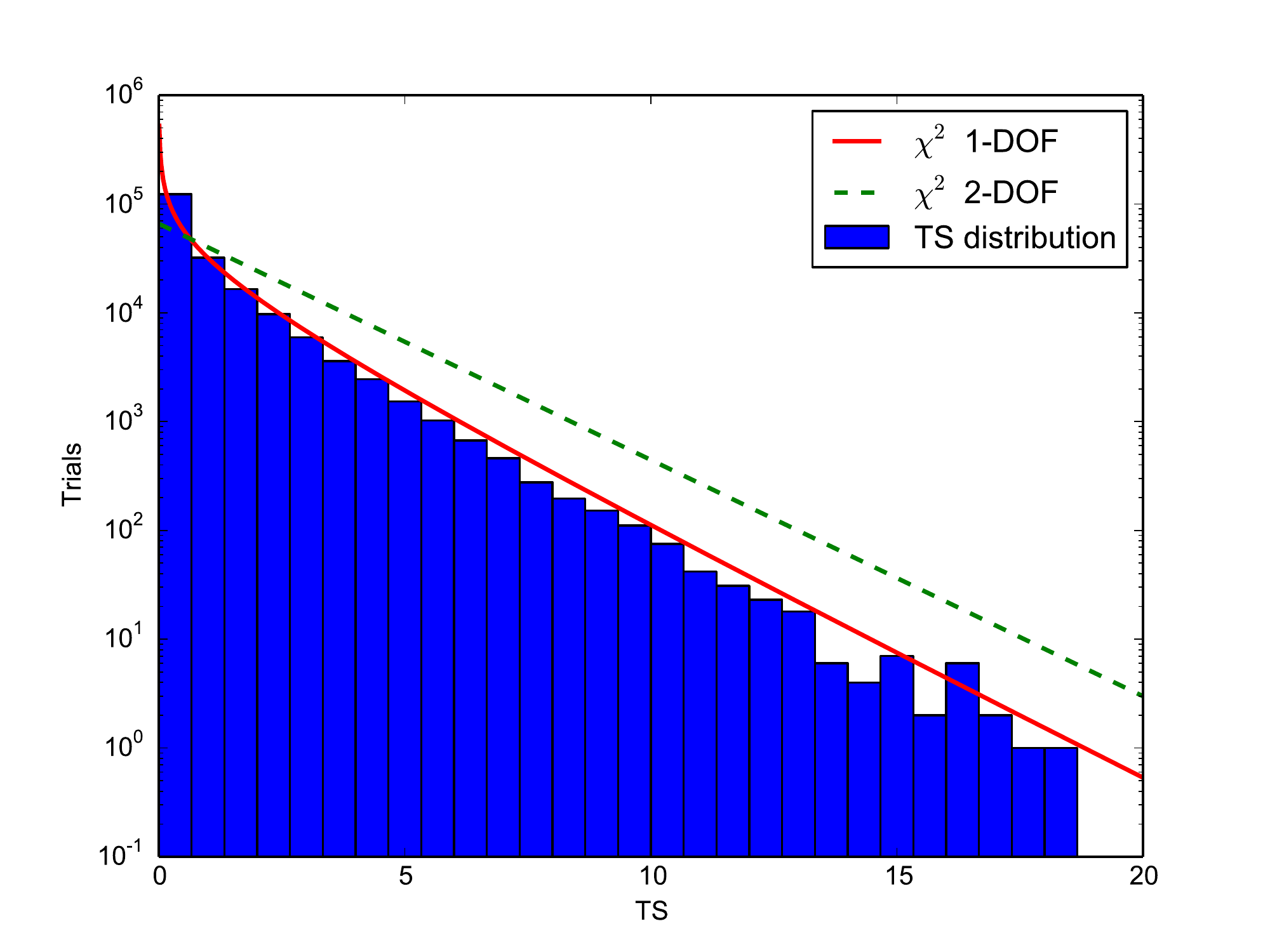} 
\caption{\label{fig:chi2} Distribution of the test statistics for ${\mathcal O}(10^7)$ number of experimental measurements of the number of events for true values of the phases $(\delta,\phi_1)=(0,0)$ for IH and $(\gamma,\theta,M_1)=(3.5,0,1)$ GeV, compared to the $\chi^2$ distribution for 1 or 2 degrees-of-freedom.}
\end{center}
\end{figure}

We have evaluated via  Monte Carlo the statistical distribution of this test statistics in order to define confidence intervals for the exclusion of the CP conserving hypothesis, following the approach of \cite{Blennow:2014sja}. In Fig.~\ref{fig:chi2} we show the result of ${\mathcal O}(10^7)$ experiments where the true value for the phases is any of the 
CP conserving hypothesis and the number of events is distributed according to Poisson statistics. The distribution is compared with a $\chi^2$ distribution of one or two degrees of freedom. The true distribution  is very similar for NH/IH and is very well approximated by the $\chi^2(1$dof). This is probably  due to the strong correlation between the two CP phases. We conclude from this exercise that is a good approximation to define the $5\sigma$ regions in the following as those corresponding to $\Delta \chi^2 = 25$. 

\begin{figure}[h]
 \begin{center}
\includegraphics[scale=0.35]{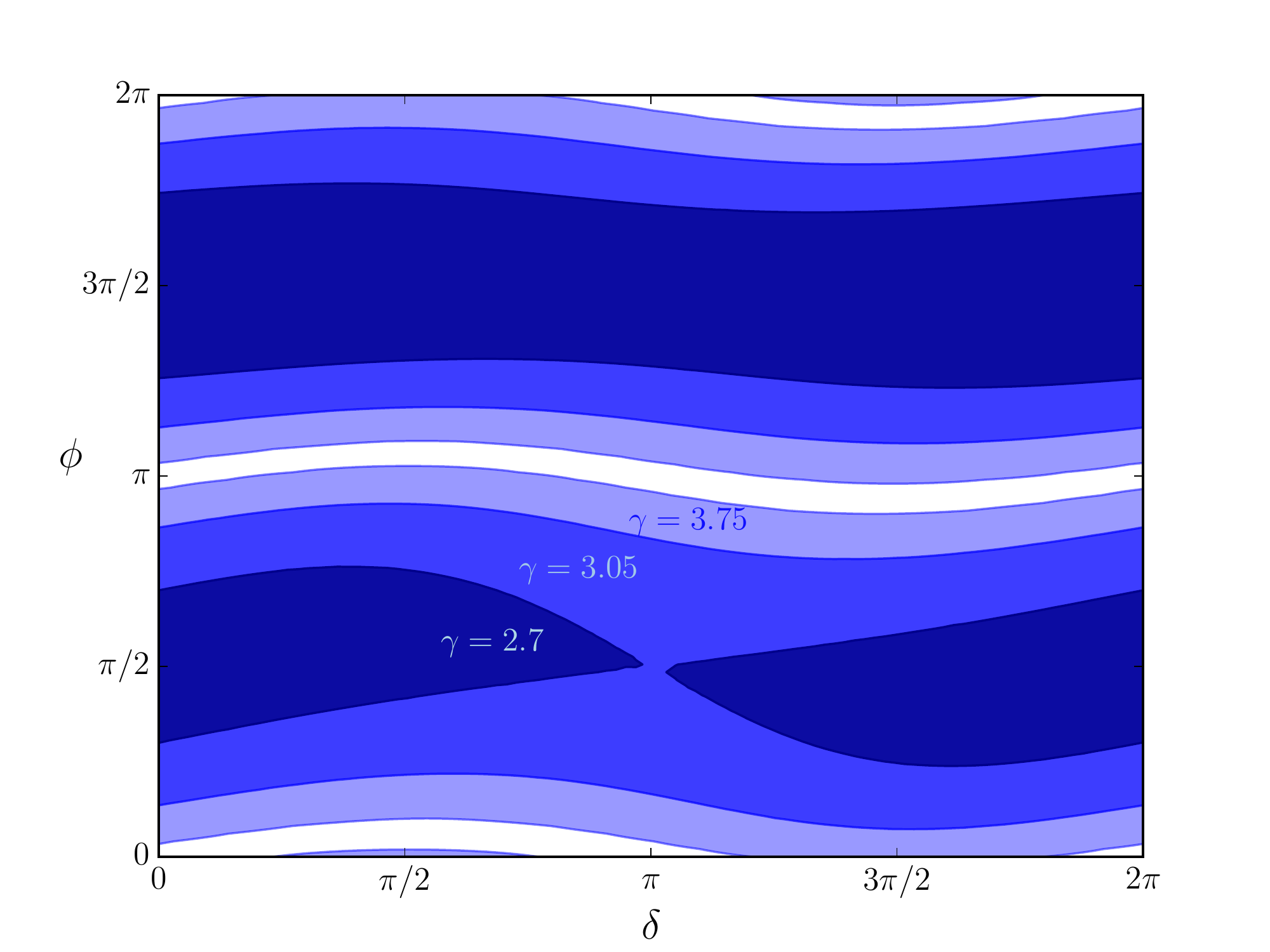} 
\includegraphics[scale=0.35]{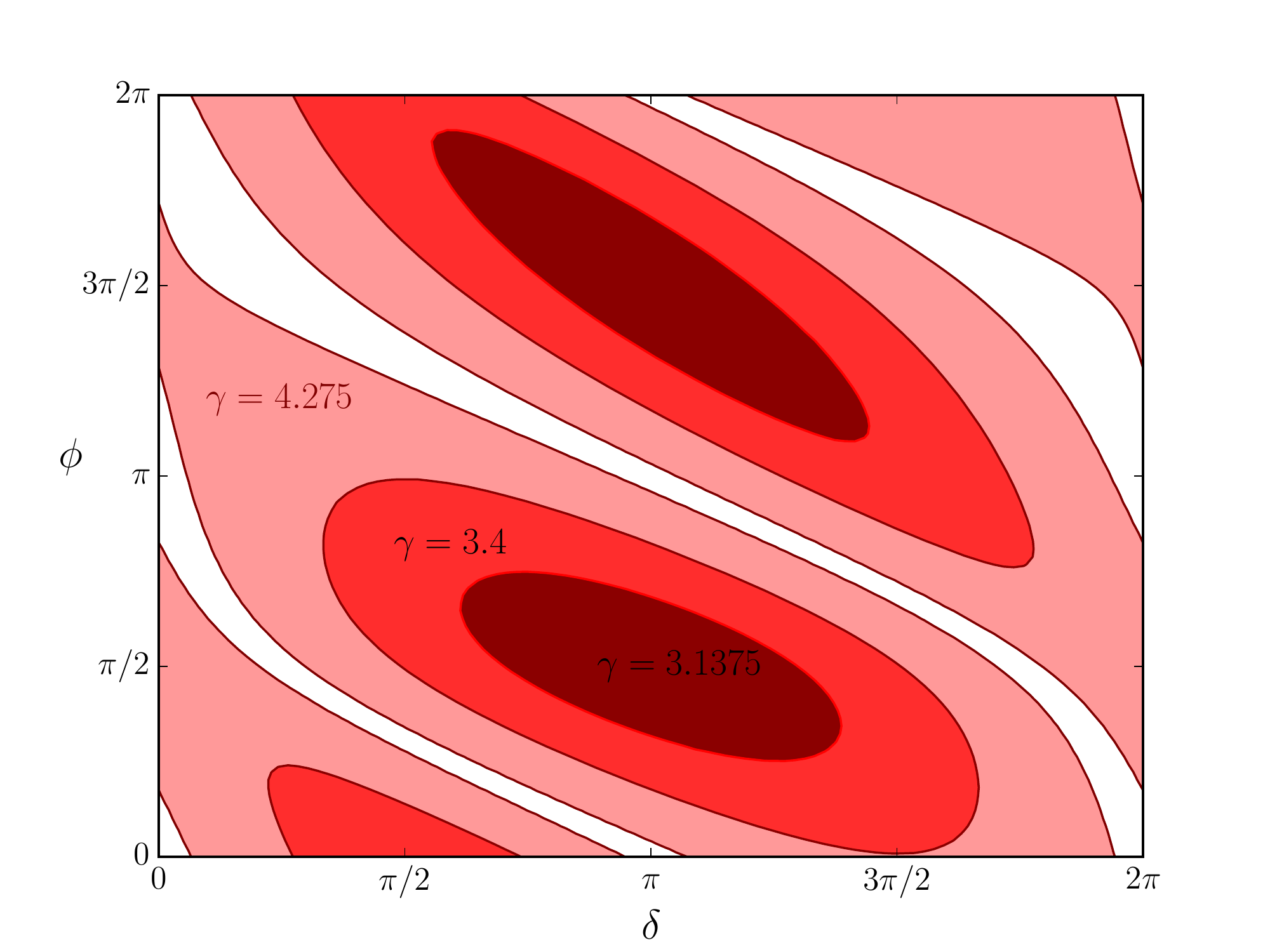} 
\caption{\label{fig:contoursgamma} $5\sigma$ CP-violation SHiP discovery regions on the plane $(\delta, \phi_1)$ for various true values of $\gamma$ for IH (up) and NH (down) and for the true parameters $M_1=1$ GeV and $\theta=0$.}
\end{center}
\end{figure}

As indicated by the analytical results, we expect almost no dependence on the parameter $\theta$, while the sensitivity to CP violation is expected to depend
significantly on $\gamma$ and $M_1$ that control the size of the mixing. 
In Fig.~\ref{fig:contoursgamma} we show the $5\sigma$ contours with $M_1=1$ GeV and for different values of true $\gamma$ on the plane $(\delta, \phi_1)$.  The coloured regions limited
by the contours indicate the range of true CP phases for which leptonic CP conservation can be excluded at $5\sigma$ CL. 

\begin{figure}[h]
 \begin{center}
\includegraphics[scale=0.35]{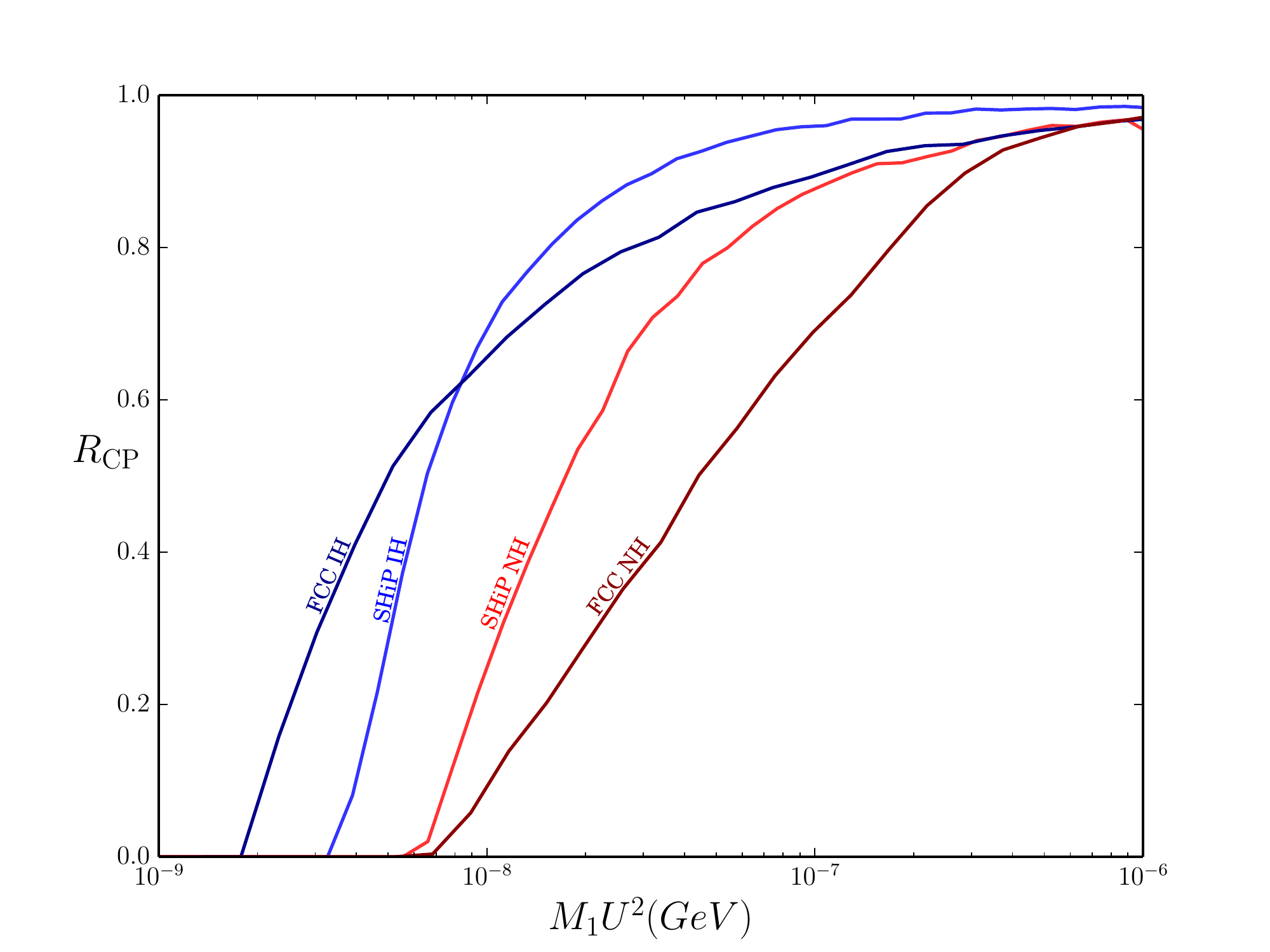} 
\includegraphics[scale=0.35]{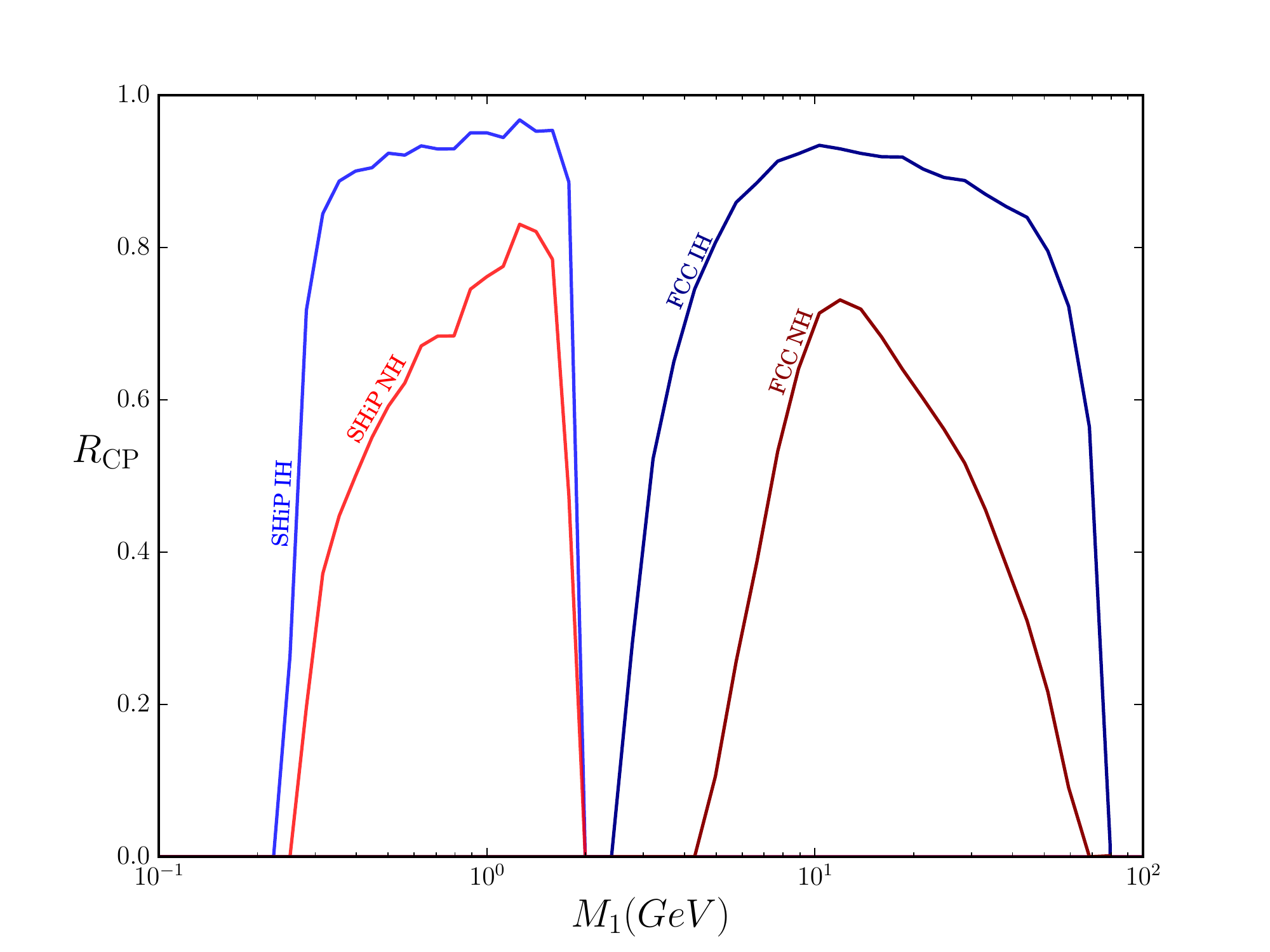} 
\caption{\label{fig:cpfraction} CP fractions for a $5\sigma$ discovery of leptonic CP violation for the two neutrino orderings and the two experiments SHiP and FCC-ee, as a function of $\gamma$ (shown in units of the more physical quantity $M_1 U^2$)  for $M_1=1$ GeV in SHiP and $M_1 = 30$ GeV in FCC-ee (up), and as a function of $M_1$ for $\gamma=4$ (down). }
\end{center}
\end{figure}

A nice way of quantifying the potential of SHiP and FCC-ee is provided by the CP fraction, $R_{CP}$,  defined as the fraction of the area of the CP phase square $\delta, \phi_1 \in [0,2 \pi]$ where leptonic CP violation can be discovered (ie. the CP conserving hypothesis excluded) at $5\sigma$ CL,  in complete
analogy as the CP fraction for the $\delta$ phase, often used in evaluating the sensitivity to CP violation of future long-baseline neutrino oscillation experiments. In  Figs.~\ref{fig:cpfraction} we show the  $R_{CP}$ as a function of $\gamma$ and $M_1$. 

Some observations are in order. CP fractions are significantly larger for IH than for NH, for the same $\gamma$. This can be understood because the dependence of the mixings on the phases is more suppressed  in the latter. 
 There is a very significant sensitivity to leptonic CP violation
in these observables (mass and mixing to electron and muon). In SHiP
with $M_1=1$ GeV, we obtain fractions above $70$\% for
$\gamma\geq 3$ for IH (mixings of $\geq\mathcal{O}(10^{-8})$) and for $\gamma\geq 3.8$ 
for NH (mixings of $\geq\mathcal{O}(3\cdot10^{-8})$). For FCC-ee, with a mass 30 GeV,
we get fractions above $70$\% for $\gamma\geq 3.1$ (mixings above $\mathcal{O}(4\cdot10^{-10})$)
for IH and for $\gamma\geq 4.5$ (mixings above $\mathcal{O}(4\cdot10^{-9})$) for NH.
The complementarity of SHiP and FCC-ee is clear, since they have sensitivity in different  mass ranges. For mixings in the middle of their sensitivity region, they can 
reach fractions above $50\%$($90\%$) for NH(IH) in the mass range $0.4-2$GeV in SHiP, and above 50$\%$ (80\%) for NH (IH) in the range $7-30$GeV in FCC-ee. 
 
Our analysis relies on rather simple assumptions, and should be further improved by including a realistic detector response and systematic errors. However the potential of this measurement is excellent and we do not expect a big difference in the results when a more realistic simulation is implemented. It will also be interesting to include the extra information provided by the measurement of the mixing to $\tau$ leptons, that should break the $\delta, \phi_1$ correlation to a
discrete degeneracy \cite{Hernandez:2016kel}.  Note also that we have only considered the contribution of one sterile state, but if there is another state 
in the same mass range, the statistics would improve by a factor of two.

\section{Conclusions}
 
 In seesaw models, the size of the mixing of the heavy Majorana neutrinos is strongly correlated with their masses, while their flavour structure is strongly dependent on the structure of the PMNS matrix. In the minimal seesaw model with just two right-handed neutrinos, this correlation is strong enough that the ratio of the mixings
 to electron and muon flavours is essentially fixed by the PMNS CP phases. In this letter we have quantified for the first time the sensitivity of these observables to the existence of leptonic CP violation. We have considered two proposed experiments SHiP and FCC-ee that have a superb sensitivity to heavy neutrinos in the mass range between ${\mathcal O}(1-100)$GeV. Within their range of sensitivity, we have demonstrated that the discovery of a massive neutrino and the measurement of its mass and its mixings to electrons and muons can result in a $5\sigma$ CL discovery of leptonic CP violation in very significant fraction of the  CP-phase parameter space
 (> 70$\%$  for mixings above ${\mathcal O}(1/3 \cdot 10^{-8})$ for IH/NH in SHiP and above  ${\mathcal O}(4 \cdot 10^{-10}/4 \cdot 10^{-9})$ in FCC-ee).

\section*{Acknowledgements}
We thank  A.~Blondel and N.~Serra useful discussions. 
This work was partially supported by grants FPA2014-57816-P, PROMETEOII/2014/050 and SEV-2014-0398,  as well as by  the EU projects 
H2020-MSCA-RISE-2015 and H2020-MSCA-ITN-2015//674896-ELUSIVES.

\end{document}